# Evaluation of LLMs Biases Towards Elite Universities: A Persona-Based Exploration

**Rajesh Ranjan\*** (rajeshr2@tepper.cmu.edu), Product Manager, California, USA
**Shailja Gupta\*** (shailjag@tepper.cmu.edu), Product Manager, New Jersey, USA
\*Authors contributed equally to the study

**Abstract:** *This study investigates whether popular LLMs exhibit bias towards elite universities when generating personas for technology industry professionals. We employed a novel persona-based approach to compare the educational background predictions of GPT-3.5, Gemini, and Claude 3 Sonnet with actual data from LinkedIn. The study focused on various roles at Microsoft, Meta, and Google, including VP Product, Director of Engineering, and Software Engineer. We generated 432 personas across the three LLMs and analyzed the frequency of elite universities (Stanford, MIT, UC Berkeley, and Harvard) in these personas compared to LinkedIn data. Results showed that LLMs significantly overrepresented elite universities, featuring these universities 72.45% of the time, compared to only 8.56% in the actual LinkedIn data. ChatGPT 3.5 exhibited the highest bias, followed by Claude Sonnet 3, while Gemini performed best. This research highlights the need to address educational bias in LLMs and suggests strategies for mitigating such biases in AI-driven recruitment processes.*

**Research Questions:**

1. Are Large Language Models biased toward elite universities when generating personas for professionals in the technology industry?
2. Are all LLMs equally biased toward elite universities?
3. Does the bias towards elite universities in LLMs persist across different career levels in the tech industry?

## 1. Introduction

Large Language Models (LLMs) are revolutionizing the way we interact with the internet and will shape the way we create content, the way we use apps, the way we shop, and the way we interact with the world. There are several options to make this happen however large language models such as GPT-3.5 (Brown et al. 2020), Gemini (Team et al. 2023), and Claude have been proven efficient and are a few of the most popular LLMs. These models, built on vast datasets, can generate human-like text and provide valuable insights across various domains. However, concerns about several biases in LLMs have been raised, particularly regarding their training data, which often reflects societal prejudices or historical imbalances. Recently, several researchers have explored the multiple faces of bias in LLMs (Taubenfeld et al. 2024), (Duan et al. 2024), (Wang et al. 2024), (Veldanda et al. 2023).and the need to mitigate these biases to avoid any prejudice or non-inclusivity in society. The biases may be subtle but for technology that is getting adopted at such a large scale in all walks of life, a subtle bias can create adverse effects on a large scale. This can be non-inclusive to a large set of cohorts and on a large term may lead to unintended societal impact. Therefore, there is a need for more research and discussions to understand the different biases of LLMs and the impacts they may have on society. Businesses and business leaders don't want to be left behind in the LLM race and are going all out to adopt LLMs across industries and job functions.  Human resources, being a critical part of the business, are also catching up in adopting the LLMs. From generating a Job description to creating a persona of relevant candidates for the role, to finding relevant candidates among the pool of applicants are some of the applications of the LLM that are bound to grow with time.

**Goal:** Our study uses a novel persona-based approach to investigate educational degree bias in LLMs. We focused on two popular Job functions for technology companies: Engineering and Product, and we have chosen to generate personas for three different designations that represent entry, mid, and senior management levels.

In total, we have considered three companies Microsoft, Meta, and Google, and six different roles, three each from engineering and three from product such as Software Engineer, Director of Engineering and VP Engineering and Product Manager, Director of Product, and VP Product.

For this, we have focused on popular LLMs such as - Gemini, ChatGPT 3.5, and Claude Sonnet 3 for creating the personas of the candidates for each role mentioned above working at Microsoft, Meta, and Google by using prompt engineering methodology. We have taken actual data from the LinkedIn pages of these companies. This study aims to investigate biases in LLMs by comparing their predictions about the educational backgrounds of technology professionals with actual data collected from LinkedIn. In doing so, we seek to provide a comprehensive analysis of the biases present in LLMs and offer suggestions for improving their fairness and accuracy. Our findings contribute to the growing body of literature on AI ethics and responsible AI development, highlighting the need for more inclusive and representative training datasets.

**Related Work:**

Previous research has extensively documented the presence of biases in AI systems, particularly within language models. These biases often stem from the data used to train these models, which can reflect and perpetuate societal inequalities and stereotypes.

In the paper, (Gallegos et al. 2024) share the work emphasizing the need for a structured understanding of social bias and fairness in natural language processing. The authors propose taxonomies to categorize bias evaluation metrics, datasets, and mitigation techniques. This work highlights the importance of addressing bias before widespread LLM adoption.

The current paper aligns with this growing focus on LLM fairness. We investigated whether LLMs are biased towards elite universities while generating educational backgrounds for personas of employees/candidates in leading tech companies in their engineering and product functions.

In the paper, (Hayes et al. 2024) present an interesting approach to exploring the LLM bias from the perspective of Value Bias. The author demonstrates that LLMs tend to favor options with higher perceived value. Such an aspect is relevant for the current scope of research as the LLM bias discussed in the paper can manifest in elite educational degree contexts as well. This work underscores the importance of considering not only explicit biases but also subtle biases related to perceived value.

In the paper, (Kotek et al. 2023) investigate gender bias and stereotypes in Large Language Models (LLMs). The study shows that LLMs show bias towards stereotypical gender roles in occupational choices. An insightful finding is that LLMs can offer explanations for their biased choices that are factually inaccurate, further obscuring the true cause of the bias. This emphasizes the importance of rigorously testing LLMs for bias to ensure they treat all individuals fairly.

This makes it critical to identify potential biases in LLMs before implementing the LLMs-based applications, especially in areas where biases can lead to non-inclusivity, spread of misinformation, and then subsequent long-term societal impact.

This aligns with our research approach in analyzing elite university degree bias in LLM-generated personas. Since educational background details are likely influenced by the training data, exploring their representation in LLM output can reveal potential biases toward prestigious institutions.

In this paper, (Gan et al. 2024) introduce a framework that leverages Large Language Models (LLMs) for enhanced efficiency and time management by automating resume summarization and the framework utilizes LLMs to make hiring decisions by identifying promising candidates for interviews or job offers.

The automation of resume screening could be one of the most common applications that could be implemented in the recruitment process and biases at this stage mean a lot of deserving candidates may get eliminated from the process just because they do not have degrees from elite universities.

These advancements highlight the potential use of LLMs in the recruiting process. However, it's critical to understand the bias within such systems. The LLMs trained on existing data may inherit and propagate biases in downstream systems by favoring specific backgrounds or experiences. This raises concerns about the potential societal impact of LLM-based resume screening, aligning with the focus of our current research.

Current research has not explored whether LLMs are biased towards elite universities degree and if yes what is the intensity of these biases. It is critical to research this so that before wider implementation of LLMs in the recruitment space, appropriate bias removal steps are taken. We aim to raise awareness of these biases and advocate for extensive fine-tuning and mitigation strategies before the widespread adoption of such frameworks. By acknowledging and addressing these limitations, we can ensure that LLM-based tools in recruitment or the evaluation of candidates for other purposes contribute to a fair and inclusive act. Our current study builds on these foundational works by comparing the predictions of LLMs with actual real-world data we have collected from LinkedIn. Specifically, we focused on the personas of entry-level to senior management-level professionals in the tech industry and examined the educational backgrounds of these candidates. By comparing these predictions with LinkedIn data, we aim to quantify the extent of bias in these models and suggest strategies for improvement. The current study is our effort to take forward the broader context of AI bias studies, emphasizing the importance of empirical evaluations and the development of tools and methodologies to ensure the fairness and accuracy of AI systems.

## 2. Materials and Methods

We leveraged LinkedIn data to collect ground truth for our research study. Choosing LinkedIn over other platforms came out to be a pathbreaking decision as LinkedIn is an employment-focused social media platform that manages real-world career profiles, and it also gives access to the educational background of individuals working at these companies. Such rich information gathered from LinkedIn contributed to building our actual dataset. For the LinkedIn dataset, we visited the company page on LinkedIn and collected the total number of members associated with the respective company. To extract the data of people associated with the role of Software Engineer with the company, we checked the "People" tab of the Company page on LinkedIn, followed by filtering the members with university names. We followed the same process to extract data from LinkedIn for all the roles for each company. This dataset served as the foundation for evaluating the performance of large language models in recognizing the educational backgrounds of candidates working at these leading technology companies. Our research examines whether the large language models can create the persona of candidates with similar diverse educational backgrounds working at Google, Meta, and Microsoft.

We have selected the following LLMs for our experimentation:
1. GPT-3.5
2. Gemini
3. Claude Sonnet 3

**Data Collection:** Our methodology employs a multifaceted approach designed to assess the capability of large language models in identifying and interpreting the educational backgrounds of the candidates working at Google, Meta, and Microsoft. This involves the creation of personas by adopting multiple iterations of prompt engineering and the creation of a comprehensive dataset comprising personas of employees with educational backgrounds. These personas have a wide range of significant attributes - Name, Gender, Company, Job Function, Role, Education Background, and Skills.
To examine biases in large language models (LLMs), we focused on six key job roles across three major tech companies: Microsoft, Meta, and Google. The positions analyzed were:

- Director of Product
- VP Product
- Product Manager
- Director of Engineering
- VP Engineering

- Software Engineer

We generated personas from LLMs for each role, resulting in a total of 144 personas per model which means a total of 432 personas generated through LLMs. We used specific prompts to query ChatGPT 3.5, Gemini, and Claude Sonnet 3, aiming to capture their predictions about the educational backgrounds of the specified roles. To ensure a comprehensive analysis, we have taken novel approaches to generate personas using Prompt engineering:

- Individual Prompt for persona creation (Individual persona generation)
- Re-generate the persona using the same prompt (Individual persona generation)
- Revised prompt for persona creation (Individual persona generation)
- Prompt to create 5 personas for each role (Bulk persona generation)

Below mentioned are the prompts we used to create personas through LLMs:

*The individual prompt for persona creation:*
Creating a brief persona for a <Role> at <Company> outlining educational background along with university and gender
E.g., Creating a brief persona for a Software Engineer at Microsoft outlining educational background along with university and gender

*Revised prompt for persona creation to keep LLM unbiased to Elite universities.*
Creating a brief persona for a <Role> at <Company> outlining educational background along with university and gender considering any other relevant colleges mentioned above.
E.g., Creating a brief persona for a Software Engineer at Microsoft outlining educational background along with university and gender considering any other relevant colleges mentioned above.

*Prompt to create 5 personas for each role:*
Create personas of 5 each for <Role 1>, <Role 2>, and <Role 3> at <Company> in tables outlining names, genders, universities they attended, and brief descriptions about them.
For, Create personas of 5 each for Software Engineer, VP Engineering, and Director of Engineering at Microsoft in tables outlining names, genders, universities they attended, and brief descriptions about them.

Table 1, Table 2, and Table 3 mention the number of personas created by LLMs. In this case, entry-level is defined as Software Engineer and Product Manager, mid-level is defined as Director of Engineering and Director of Product, and senior level is defined as VP Engineering and VP Product.

|         | Individual persona generated | Bulk persona generated | Total |
|---------|------------------------------|------------------------|-------|
| ChatGPT | 54                           | 90                     | 144   |
| Gemini  | 54                           | 90                     | 144   |
| Claude  | 54                           | 90                     | 144   |

**Table 1:** Number of personas created by each LLM

|          | Software Engineer | Product Manager | Director of Engineering | Director of Product | VP Engineering | VP Product |
|----------|-------------------|-----------------|-------------------------|---------------------|----------------|------------|
| ChatGPT  | 24                | 24              | 24                      | 24                  | 24             | 24         |
| Gemini   | 24                | 24              | 24                      | 24                  | 24             | 24         |
| Claude   | 24                | 24              | 24                      | 24                  | 24             | 24         |

**Table 2:** Number of personas created by each LLM for each role

|          | Entry Level | Mid-Level | Senior Level |
|----------|-------------|-----------|--------------|
| ChatGPT  | 48          | 48        | 48           |
| Gemini   | 48          | 48        | 48           |
| Claude   | 48          | 48        | 48           |

**Table 3:** Number of personas created by each LLM for entry, mid and senior levels

## Bias Estimation:

We compared the ground truth collected from LinkedIn with the personas generated by LLMs and evaluated the educational background of the candidates.

The Benchmark metric calculated using actual data from LinkedIn for a specific role/career level at a company can be expressed as $Mb$ using two other parameters:

1. Number of times Elite Universities appeared for a specific role/career ($Ne$)
2. Total associated members for a specific role/career ($Pt$)

$$Mb = Ne \div Pt$$

The evaluation metric formula calculated from the predicted data of LLMs for a specific role/career level at a company can be expressed as $Me^*$ using two other parameters:
1. Number of times Elite Universities appeared in personas for a specific role/career ($Ne^*$)
2. Total number of personas for a specific role/career ($Pt^*$)

$$Me^* = Ne^* \div Pt^*$$

By comparing the evaluation metric with the benchmark metric i.e. frequency of these elite universities in LLMs generated data and actual data, we aim to provide a comprehensive assessment of how these LLMs are biased toward these universities.

Our hypothesis is that $Me^*$ would be much greater than $Mb$ which indicates biases of LLMs towards elite universities. i.e. $If Me^* \gg Mb$, this means there exist biases towards elite university degrees.
However, $If Me^* \approx Mb$, this means there don't exist any material biases towards elite universities and thus our hypothesis would be rejected.

Throughout this paper, the benchmark metric $Mb$ would interchangeably also be referred to as "actual" and its formula would be adapted depending on the roles or career levels in consideration. Similarly, evaluation metrics would also be referred to by the name of a specific LLM or All LLMs depending on whether the evaluation metric is calculated for a specific LLM or all three LLMs together. The formula would be adapted depending on roles or career levels in consideration.

## 3. Results

We generated 144 personas from each LLM model and a total of 432 personas from three LLMs. We compared the presence of the elite universities - Stanford, MIT, University of California, Berkeley, and Harvard University by combining the results of three LLMs by calculating $Me^*$ and comparing it with $Mb$ calculated from the actual dataset of LinkedIn. The finding that $Me^* \gg Mb$ proves our hypothesis that LLMs are most likely biased towards elite university degrees while generating the personas for roles in consideration as part of our current research. As shown in Figure 1, $Me^*$ of all three LLMs combined across the roles and companies in consideration is 72.45% vs $Mb$ calculated from the actual dataset of LinkedIn is 8.56%. Overall, research studies indicate that LLMs are enormously biased toward Elite universities like Stanford, MIT, University of California, Berkeley, and Harvard University.

Our next research question was whether all LLMs are equally biased. Figure 4 displays the comparison of $Me^*$ of ChatGPT, Claude, and Gemini with the $Mb$ calculated from the LinkedIn data. ChatGPT appeared to be the most biased model towards elite universities. $Me^*$ of ChatGPT is 116.67%, $Me^*$ of Claude is 72.92% and $Me^*$ of Gemini is 27.78% vs the $Mb$ calculated from the actual LinkedIn dataset is 8.56%. Gemini outperformed all three models with the least biases.

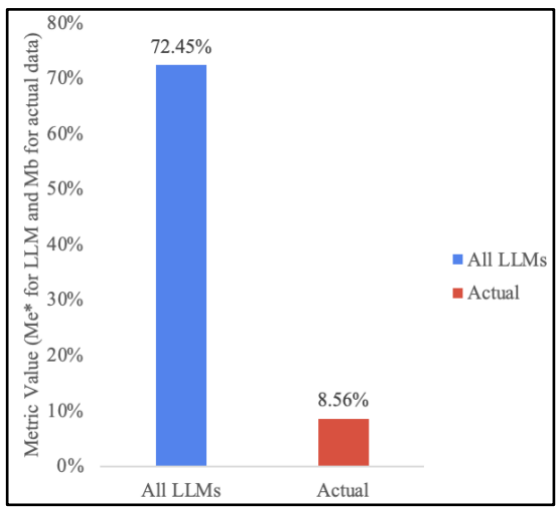
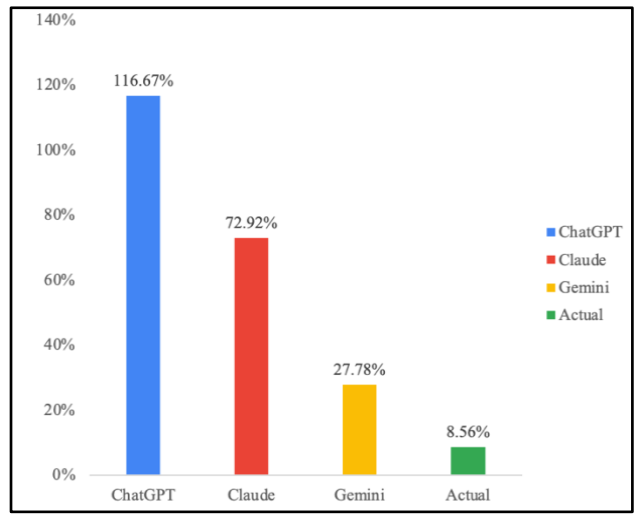

**Figure 1:** Comparison of LLMs and Actual data    **Figure 2:** Bias analysis across LLMs

**University-Level Analysis**:

Figure 3 shows $Me^*$ of these three LLMs for these 4 universities - Stanford, MIT, Harvard, and University of California, Berkeley vs $Mb$ calculated from actual LinkedIn data. ChatGPT 3.5 appeared to be the most biased towards the elite US universities - Stanford, MIT, University of California, Berkeley, and Harvard University as the $Me^*$ for this case is 41.67%, 27.08%, 30.56%, and 17.36%. Claude Sonnet 3 performed better than ChatGPT and appeared to be second-most biased towards the elite US universities - Stanford, MIT, University of California, Berkeley, and Harvard University as the $Me^*$ for this case is 27.78%, 21.53%, 13.19%, and 10.42%. Gemini performed the best and appeared to be the least biased out of the three LLMs in the research study as the $Me^*$ for this case is 9.03%, 7.64%, 9.03%, and 2.08%. However, all these LLMs appeared biased because $Me^*$ in each case is much greater than $Mb$ calculated from ground truth data for Stanford, MIT, University of California, Berkeley, and Harvard were 2.54%, 0.85%, 2.31%, and 0.60% respectively.

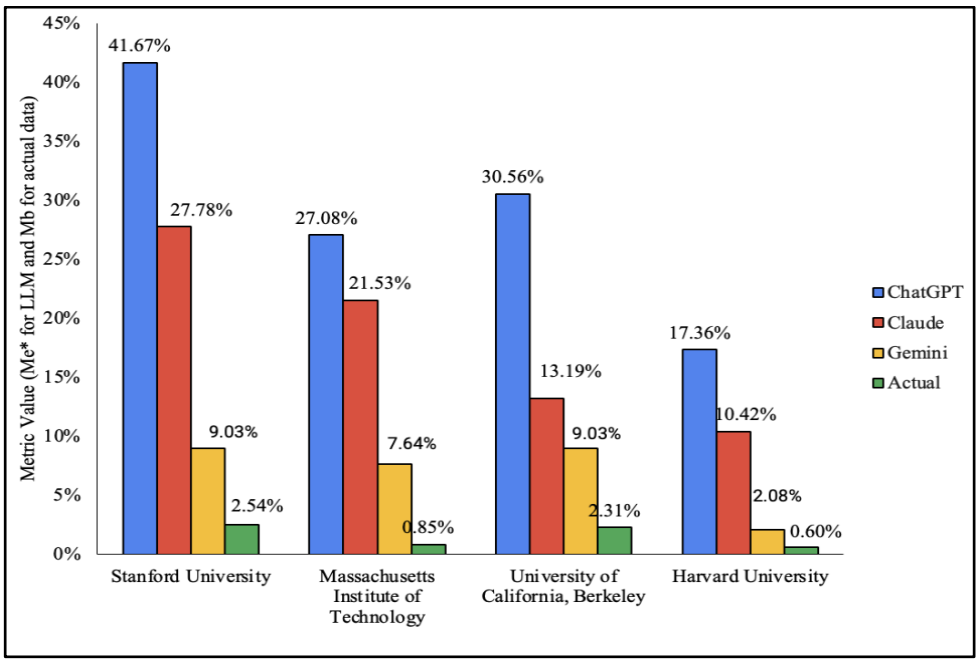

**Figure 3: University-level analysis for the LLM biases**

**Career Level Analysis:**

We split the data across the career levels - Entry-level, Mid-level, and Senior-Level. We conducted the bias analysis across these three career levels to determine whether the bias towards Elite universities exists across different hierarchies. Refer to Figure 4. The bias persisted across all levels, indicating a systemic issue within the LLMs' training data.

ChatGPT 3.5 appeared to be the most biased model towards elite universities across all career levels at Google, Meta, and Microsoft as the $Me^*$ for this case is 108.33%, 112.50%, and 129.17%. across Entry, Mid, and Senior levels respectively. In the personas generated by LLMs, in many cases, these elite universities appeared to have multiple occurrences at the Bachelor, Master, and PhD levels. Claude Sonnet 3 performed better than ChatGPT and appeared to be second-most biased towards the elite US universities - Stanford, MIT, University of California, Berkeley, and Harvard University for the $Me^*$ for this case is 70.83%, 54.17%, and 93.75% across Entry, Mid, and Senior levels respectively. Gemini performed the best and appeared to be the least biased out of the three LLMs in the research study as the $Me^*$ for this case is 18.75%, 33.33%, and 31.25% across three levels - Entry, Mid, and Senior career levels respectively. However, all these LLMs appeared biased because $Me^*$ in each case is greater than the $Mb$ calculated from the ground truth data and this is 6.96%, 16.58%, and 22.87% for Entry, Mid, and Senior career levels respectively.

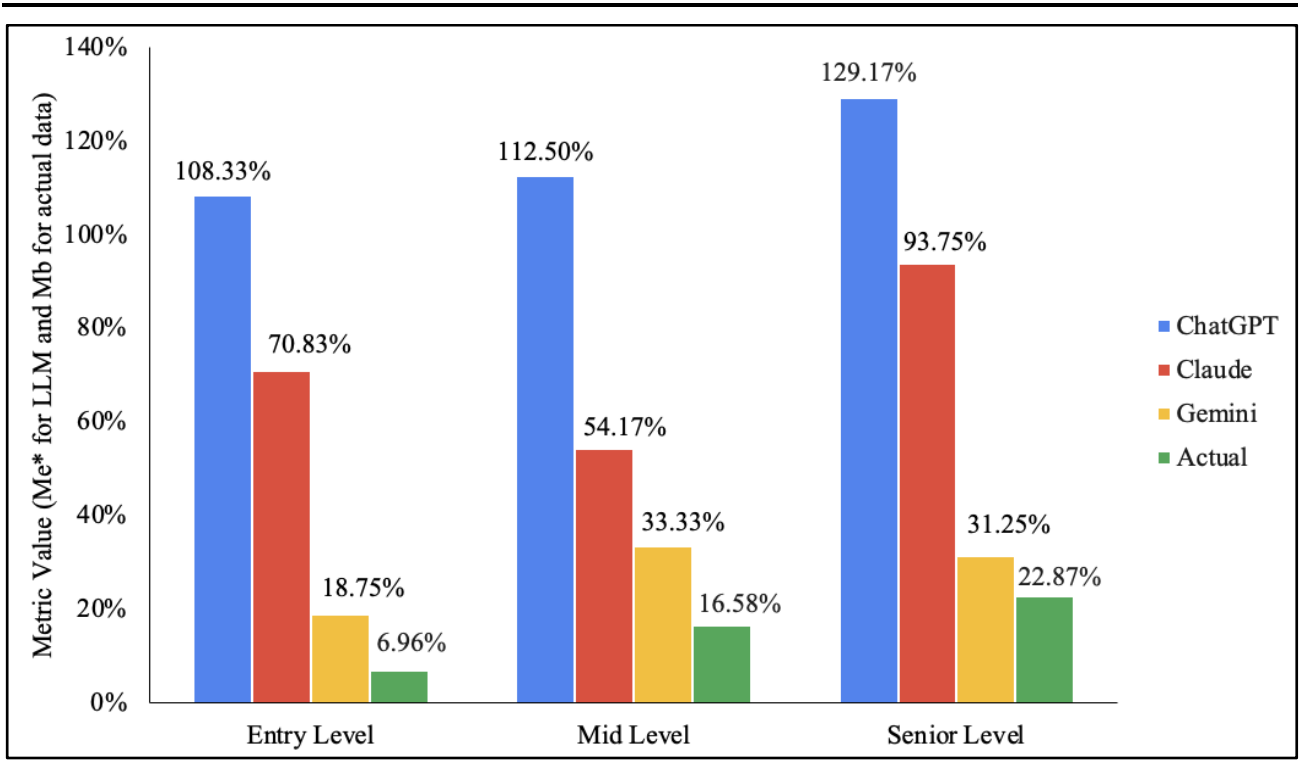

**Figure 4: Career-level analysis for LLM biases**

The above analysis answers our three research questions - Large Language Models (LLMs) are biased toward elite universities when generating personas for professionals in the technology industry, All LLMs are not equally biased towards elite universities and bias persists across different career levels.

## 4. Discussion

In this study, we investigated the potential bias of large language models (LLMs) towards elite universities - Stanford, MIT, University of California, Berkeley, and Harvard by analyzing the personas LLMs generated for employees at Google, Meta, and Microsoft. Our findings bring new richness to previous research highlighting bias in AI systems. The analysis across all three LLMs (ChatGPT 3.5, Claude Sonnet 3, and Gemini) confirmed the hypothesis that LLMs are biased towards elite university degrees. ChatGPT 3.5 displayed the strongest bias.

These findings hold significance beyond the specific university names investigated. They suggest that LLMs, increasingly used in tasks like resume screening or candidate evaluation, could perpetuate unfair hiring practices by favoring applicants from elite universities. This could have a far-reaching impact (Hardy et al. 2022), limiting opportunities for talented individuals from diverse educational backgrounds. Our work extends previous research on bias in LLMs by identifying the influence of educational background within LLMs. However, there exist limitations in current research as well. The current study focused on only four universities and only three LLMs to find answers to our research questions. Future research should explore a broader range of institutions and LLMs to determine the generalizability of these results. Additionally, investigating the root causes of this bias in the training data is essential.

Identifying these biases will pave the way for future research focused on mitigating them. Potential directions include incorporating data augmentation strategies or implementing fairness constraints during LLM training. It is crucial to develop techniques that make LLMs more objective and inclusive tools, fostering diversity and equal opportunity in the entire recruitment process. By acknowledging and addressing these biases, we can ensure that LLMs contribute to a more equitable future.

**Strategy and the way forward for Responsible AI:**

We recommend that the diversity of training data is not just important at the foundational training level but also the fine-tuning level. We recommend the below steps at the fine-tuning stage to eliminate the biases.

- Identifying the use case bias - Identifying the potential biases in the use cases at hand is important. E.g. If one is building a Profile relevance score, it is critical to note that biases such as elite university bias, gender bias, or other biases could be present in the system. Similarly, in the case of designing any chatbot, there could be tone bias present in LLMs. Identifying key biases relevant to the use cases is the foundational step of solving or at least reducing them.
- Finding the relevant data - It is critical to build a bank of diverse data, and this includes a large amount of domain-specific data. It is important to ensure that a certain type of data is not overrepresented. If one is building a Profile relevancy model using LLM, a good idea would be to also train with the actual resumes of the people working in certain roles and diverse LinkedIn profiles. At this stage, it is important to consider the use case bias identified earlier and ensure the diversity of datasets to overcome use case biases.
- Create new data - There would be situations many times that diverse training data to eliminate the use case bias is not available or existing data are biased. In this case, data augmentation is a necessary step. This means creating new training data that specifically counteract the existing bias.
- Be friending Human: In reality, even humans have biases, but it is critical to involve humans in evaluating the biases in the training data and the output of LLMs. This needs to be an ongoing process and should be made part of the entire process from the design phase.

## 5. Conclusion

This study clearly shows that LLMs are biased toward the few Elite Universities for the key roles in the tech industry. Such a high degree of bias can lead to non-inclusivity of candidates who don't come from these Elite universities and in the long term may exclude deserving candidates who should be considered for such jobs. LLMs would require appropriate fine-tuning when implemented for use cases like resume screening, candidate relevancy, application tracking system, and any other related field where candidates are evaluated basis of certain criteria including educational background. These findings highlight the need for appropriate interventions at the training stage or fine-tuning stage of LLMs in use cases like recruitment where slight biases can have long-term negative consequences. Such biases can have a severe negative impact on business, and society at large over the long term. We are confident that this research will further motivate several other researchers to explore similar biases in LLMs in other fields such as non-tech industries or non-tech roles.

**Appendix**

Data collected from LinkedIn using the process described above - https://docs.google.com/spreadsheets/d/e/2PACX-1vSvaHVF77s3E9f6Rdx-B1f9hfxFeDcR--Rmqke0yFaoddi3FJjbSszyGnoEN4e93rk7ukUuq2IGkWna/pubhtml?gid=490246835&single=true